\def\be{\begin{equation}}
\def\ee{\end{equation}}
\def\ba{\begin{array}}
\def\ea{\end{array}}
\def\t{\tilde}
\def\L{\Lambda}
\def\l{\lambda}
\def\p{\prime}
\def\pp{{\prime\prime}}
\def\Rb{{I\!\! R}}
\def\Zb{{Z\!\!\! Z}}

\def\Cb{\ \hbox{\vrule width 0.6pt height 7pt depth 0pt
		      \hskip -3.5 pt} C}
\documentstyle[12pt]{article}
\begin{document}
\parskip=4pt
\parindent=18pt
\baselineskip=22pt
\setcounter{page}{1}
\centerline{\Large\bf Integration Approach to Ising Models}
\vspace{6ex}
\centerline{\large{\sf Sergio Albeverio$^\star$} ~~~and~~~ {\sf Shao-Ming Fei}
\footnote{\sf On leave from Institute of Physics, Chinese Academy of Sciences,
Beijing}}
\vspace{4ex}
\parindent=40pt
{\sf Institute of Mathematics, Ruhr-University Bochum,
D-44780 Bochum, Germany}\par
\parindent=35pt
{\sf $^\star$SFB 237 (Essen-Bochum-D\"usseldorf); 
BiBoS (Bielefeld-Bochum);\par
\parindent=40pt
CERFIM Locarno (Switzerland)}\par
\vspace{6.5ex}
\parindent=18pt
\parskip=6pt
\begin{center}
\begin{minipage}{5in}
\vspace{3ex}
\centerline{\large Abstract}
\vspace{4ex}
An integral representation of the partition function
for general $n$-dimensional Ising models with nearest or non-nearest
neighbours interactions is given. The representation is used to derive 
some properties of the partition function. An exact solution is given in
a particular case.
\end{minipage}
\end{center}

\newpage

As a simple prototype of a statistical mechanical system that undergoes
a phase transition for spatial dimensionality $n>1$, the Ising(-Lenz) model
\cite{ising} has
been extensively studied in various ways. Since the exact
solution of the free energy $F$ and the spontaneous magnetization
for the two dimensional zero-field Ising model (on a square lattice) were
obtained more than
fifty years ago \cite{onsager,yang}, many efforts have
been made towards a detailed study of properties and the possible
finding of exact solutions for the higher dimensional Ising model or for
a two-dimensional Ising model with nonzero magnetic field, for reviews
see e.g., [4-7].
In this paper we study the Ising models by an ``integration approach". 
We present an integral representation of the partition function
for general $n$-dimensional Ising models with either nearest or non-nearest
neighbours interactions. From this representation we
prove that the partition function of n-dimensional homogeneous (i.e. translation
invariant) Ising systems
on a square lattice is uniquely given by the eigenvalues of the
related interaction coupling matrix. 
For one-dimensional homogeneous ferromagnetic systems
with positive coupling coefficients the partition function 
satisfies a special equality which means that the variation of the interaction
couplings of the system is related to the variations of the external
magnetic field and the temperature. 
For some special cases of interaction couplings, in all dimensions,
we get an exact
representation of the partition function in terms of a Bessel function.
We also give an approximate representation of 
the general partition function for the homogeneous Ising
model, which shows that its leading term is given by the largest
positive eigenvalue
(resp. the largest absolute value of the negative eigenvalues)
of the related interaction coupling matrix in the ferromagnetic (resp.
antiferromagnetic) case. 
In the case of nearest neighbours interactions in two dimensions our
integral representation is connected with the study of the Ising limit of
the $\phi^4_2$ double-well models, see e.g. \cite{dw}.

The partition function of the Ising models with the usual nearest neighbours
interactions on a $d$-dimensional lattice
with $\Lambda$ lattice sites is given by, see e.g. \cite{baxter},
\be\label{1}
Z^0_\L=\displaystyle\sum_{\{\sigma_i\}}exp\left\{
K\displaystyle\sum_{<ij>_0}\sigma_i\sigma_j
+K^\p\displaystyle\sum_{i=1}^\Lambda\sigma_i\right\},
\ee
where $\sigma_i$, $i=1,2,...,\Lambda$, takes values $+1$ and $-1$, $<ij>_0$
denotes the nearest-neighbour pairs on the lattice and $K=J/kT$,
$K^\p=H/kT$ with $T$ the temperature, $k$ the Boltzmann constant, $H$
the external magnetic field and $J$ the coupling constant. $J$ is
positive for ferromagnetic systems and negative for
antiferromagnetic systems.

We consider the partition function (\ref{1}) in a more general form
including both nearest and non-nearest neighbours interactions,
\be\label{2}
Z_\Lambda=\displaystyle\sum_{\{\sigma_i\}}exp\left\{K
\displaystyle\sum_{i,j=1}^\Lambda C^{0}_{ij}\sigma_i\sigma_j
+K^\p\displaystyle\sum_{i=1}^\Lambda\sigma_i\right\},
\ee
where $C^0 =(C^0_{ij})$ is a symmetric $\L\times\L$ matrix,
\be\label{3}
C_{ii}^0=0,~~~~C_{ij}^0=C_{ji}^0,~~~i\neq j,~~~i,j=1,2,...,\L.
\ee
$C_{ij}^0$ stands for the interaction coupling between $\sigma_i$ and $\sigma_j$.
Clearly if $C^0_{<ij>_0}=1$ and $C^0_{ij}=0$ 
for all non-nearest neighbours
pairs, the partition function (\ref{2}) is reduced to (\ref{1}). In the
following we call $C^0$ the interaction coupling matrix.

Let $<ij>_1$, $<ij>_2$ and $<ij>_\alpha$ denote the next-to-nearest
neighbours, next to next-to-nearest neighbours and $\alpha$-th 
next-to-nearest neighbours interactions, respectively.
In the following we consider systems such
that all interaction coupling constants for any given $\alpha$
are the same, i.e., $C_{<ij>_\alpha}\equiv c_\alpha$, $\forall
i,j=1,2,...,\L$, $\alpha=0,1,2,...,\L-1$. We also assume that the number
of sites interacting with an arbitrary given site $i$ is the same for all
$i=1,...,\L$. We call these systems homogeneous. In particular this
implies that we impose the periodic boundary condition on the boundary of
the lattice if the lattice is a bounded portion of $\Zb^d$
(or equivalently the lattice is identified with a $d$-dimensional torus).

Let $l_i$ be the number of lines connected to a lattice 
site $i$. $l_i$ is then a topological invariant of the lattice
and is called the topological link number associated with the lattice site $i$.
For homogeneous systems with only nearest-neighbour interactions,
the link number $l_i$
is exactly equal to the number of the sites interacting with the
lattice site $i$, and by homogeneity we have that $l_i$ is independent 
of $i$, i.e., $l_i=l$ for all $i=1,...,\L$. For instance, for a periodic 
one-dimensional chain with nearest-neighbour interactions we have $l=2$.
For a two (resp. three) dimensional square lattice with
nearest-neighbour interactions and periodic boundary
conditions we have $l=4$ (resp. $l=6$). Therefore for a homogeneous system 
in $d$ dimensions with only
nearest-neighbour interactions (i.e., such that 
$C^0_{<ij>_0}=1$ and $C^0_{ij}=0$ for all 
non-nearest neighbours pairs), the matrix $C^0$ has the following properties:
\be\label{4}
\displaystyle\sum_{i=1}^\L C_{ij}^0=\displaystyle\sum_{i=1}^\L C_{ji}^0=l,
~~~\forall j=1,2,...,\L;
\ee
\be\label{4h}
\displaystyle\sum_{i,j=1}^\L C_{ij}^0=\L l;~~~~Tr(C^0)=0.
\ee

In fact, a higher dimensional system with nearest-neighbour interactions
can always be viewed as a lower dimensional system with
non-nearest-neighbour interactions, and vice versa. For example, for a
one dimensional chain with nearest and next-to-nearest neighbours interactions,
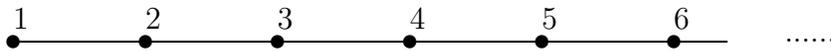
\parskip=0pt
\parindent=0pt

\begin{center}
\bigskip
\begin{picture}(300,40)(-10,-10)
\put(0,0){\line(2,0){270}}
\put(0,0){\circle*{5}}\put(50,0){\circle*{5}}\put(100,0){\circle*{5}}
\put(150,0){\circle*{5}}
\put(200,0){\circle*{5}}
\put(250,0){\circle*{5}}
\put(0,5){1}\put(50,5){2}\put(100,5){3}
\put(150,5){4}\put(200,5){5}\put(250,5){6}\put(280,0){~~~......}
\end{picture}
\bigskip
\end{center}

we have $C_{i,i+1}^0\neq 0$, $C_{i,i+2}^0\neq 0$, i.e., a lattice site
$i$ has interactions with the lattice sites $i\pm 1$ and $i\pm 2$. This
system is equivalent to a two dimensional band with nearest-neighbour
interactions,

\bigskip
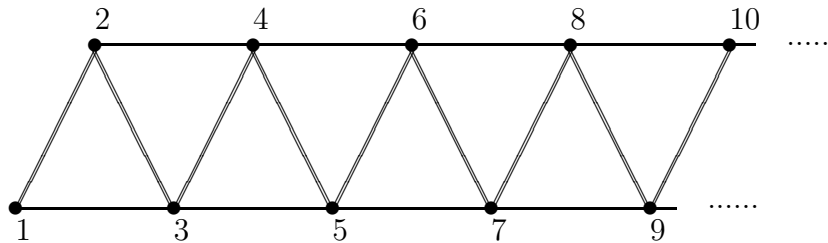
\begin{center}
\begin{picture}(300,80)(-10,-10)
\put(30,62){\line(2,0){250}}
\put(0,0){\line(2,0){250}}
\put(280,62){~~~......}
\put(250,0){~~~......}
\put(30,60){\line(1,-2){30}}
\put(90,60){\line(1,-2){30}}
\put(150,60){\line(1,-2){30}}
\put(210,60){\line(1,-2){30}}
\put(1,0){\line(1,2){30}}
\put(61,0){\line(1,2){30}}
\put(121,0){\line(1,2){30}}
\put(181,0){\line(1,2){30}}
\put(241,0){\line(1,2){30}}
\put(0,0){\line(1,2){30}}
\put(60,0){\line(1,2){30}}
\put(120,0){\line(1,2){30}}
\put(180,0){\line(1,2){30}}
\put(240,0){\line(1,2){30}}
\put(29,60){\line(1,-2){30}}
\put(89,60){\line(1,-2){30}}
\put(149,60){\line(1,-2){30}}
\put(209,60){\line(1,-2){30}}
\put(0,0){\circle*{5}}\put(60,0){\circle*{5}}\put(120,0){\circle*{5}}
\put(180,0){\circle*{5}}\put(240,0){\circle*{5}}
\put(30,62){\circle*{5}}\put(90,62){\circle*{5}}\put(150,62){\circle*{5}}
\put(210,62){\circle*{5}}\put(270,62){\circle*{5}}
\put(0,-12){1}\put(60,-12){3}\put(120,-12){5}\put(180,-12){7}\put(240,-12){9}
\put(30,68){2}\put(90,68){4}\put(150,68){6}\put(210,68){8}\put(270,68){10}
\end{picture}
\end{center}

where the double lines stand for nearest-neighbour interactions in the one
dimensional chain and the single lines stand for next-to-nearest 
neighbours interactions in the one dimensional chain. 
If different interaction couplings in different
directions are assumed then we have $C_{i,i+1}^0\neq C_{i,i+2}^0$.
\parskip=3pt
\parindent=18pt

To describe the interaction properties for both systems with nearest and
non-nearest-neighbour interactions, we define
a generalized link number $L$ of the lattice by,
\be\label{link}
L\equiv\displaystyle\sum_{i=1}^{\L}C_{ij}^0=
\displaystyle\sum_{i=1}^{\L}C_{ji}^0,~~~\forall
j=1,2,...,\L.
\ee
When the systems have only 
nearest-neighbour interactions $L$ is equal to the topological
link number $l$ of the lattice. Instead of (\ref{4h}) we have, for
homogeneous systems with either nearest-neighbour or 
non-nearest-neighbour interactions (in arbitrary dimension),
\be\label{ll2}
\displaystyle\sum_{i,j=1}^\L C_{ij}^0=\L L;~~~~Tr(C^0)=0.
\ee

Any real symmetric matrix $M$ (over the complex numbers $\Cb$) 
can be diagonalized by an orthogonal similarity
transformation. The matrix used to diagonalize $M$ has as its
columns an orthonormal set of eigenvectors for $M$. The resulting
diagonal matrix has as its diagonal elements the eigenvalues of $M$. 
Let $u_i$ be an orthonormal basis in $\Cb^\L$ (as a $\L$-dimensional Hilbert
space, with scalar product $\cdot$) of the (column)
eigenvectors of $C^0$, with eigenvalues $\l_i^0$, $i=1,2,...,\L$,
\be\label{c1}
C^0 u_i=\l^0_i u_i,~~~~\t{u}_i\cdot u_j =\delta_{ij},~~~~i,j=1,2,...,\L,
\ee
(with $\t{u}_i$ the adjoint vector to $u_i$).
Let $A$ be the orthogonal matrix that diagonalizes $C^0$. Then
\be\label{c2}
A=(u_1,u_2,...,u_\L),
\ee
\be\label{c3}
\t{A}C^0 A=diag\{\l_1^0,\l_2^0,...,\l_\L^0\},
\ee
where $diag\{\l_1^0,\l_2^0,...,\l_\L^0\}$ denotes the $\L\times\L$
matrix having $\l_1^0,\l_2^0,...,\l_\L^0$ on the diagonal and $0$
elsewhere, and $\t{A}$ is the adjoint of $A$.
Moreover we have
\be\label{c4}
\t{A}A=1,~~~~detA=det\t{A}=1.
\ee

From (\ref{c1}) we also have
\be\label{c5}
\displaystyle\sum_{j=1}^\L(C^0)_{ij}A_{jk}=\l^0_k A_{ik}.
\ee
Summing over the index $i$ on both sides of the equation (\ref{c5}) 
and using (\ref{link}) we get
$$
(L-\l^0_k)\displaystyle\sum_{i=1}^\L A_{ik}=0.
$$
Therefore if $\l_k^0\neq L$ for some $k=1,...,\L$, then:
\be\label{c6}
\displaystyle\sum_{i=1}^\L A_{ik}=0.
\ee

Let us now consider the possibility that $\l_k^0=L$ for some $k$, i.e.,
$L$ is an eigenvalue of $C^0$. We study the eigenvector
$x=$column$(x_1,x_2,...,x_\L)\in\Cb^\L$ to the eigenvalue $L$ (if it
exists) of $C^0$. We have
$\displaystyle\sum_{j=1}^\L C^0_{ij}x_j=L x_i$, $i=1,2,...,\L$. From
(\ref{link}) we see that 
\be\label{u}
u_1\equiv\displaystyle\frac{1}{\sqrt{\L}}
\left(\ba{l}1\\[1mm]1\\[1mm]\vdots\\[1mm]1\ea\right)
\ee
is an eigenvector of $C^0$ with eigenvalue $L$ and with unit
length (up to a global phase factor).
By the orthonormality of the $u_i$ we have $\t{u}_1u_i=0$, 
$i=2,3,...,\L$. Hence from (\ref{u}) the matrix that diagonalizes $C^0$
has the properties
\be\label{c10}
\displaystyle\sum_{i=1}^\L A_{ik}=\left\{\ba{l}0,~~~k\neq
1\\[3mm]\sqrt{\L},~~~~k=1.\ea\right.
\ee

Generally the matrix $C^0$ has both positive and negative eigenvalues.
Let $\l^+_{max}$ be the largest positive eigenvalue 
and $\l^-_{max}$ be the largest absolute value of the negative eigenvalues. 
Set $\l^\pm=\pm\l^\pm_{max}\pm\epsilon$ for any small constant $\epsilon>0$. 
We set
\be\label{cp}
C^+\equiv\l^+ I-C^0
\ee
and
\be\label{cm}
C^-\equiv C^0-\l^-I,
\ee
where $I$ is the $\L\times\L$ identity matrix. 
$C^\pm$ are then positive
definite matrices with the properties:
\be\label{6}
C_{ii}^{\pm}={\pm}\l^{\pm},~~~~C_{ij}^{\pm}=C_{ji}^{\pm},~~~i\neq
j,~~~i,j=1,2,...,\L,~~~Tr(C^{\pm})={\pm}\l^{\pm}\L
\ee
and
\be\label{7}
\displaystyle\sum_{i=1}^\L C_{ij}^{\pm}=
\displaystyle\sum_{i=1}^\L C_{ji}^{\pm}={\pm}\l^{\pm}\mp L,
~~~\forall j=1,2,...,\L,
\ee
\be\label{8}
\displaystyle\sum_{i,j=1}^\L C_{ij}^{\pm}=({\pm}\l^{\pm}\mp L)\L.
\ee

The eigenvectors of $C^0$ are also eigenvectors of $C^{\pm}$,
\be\label{c7}
C^{\pm} u_i=\l_i^{\pm} u_i,~~~~with~~~~~
\l_i^{\pm}\equiv{\pm}\l^{\pm}\mp\l_i^0,~~~~i=1,2,...,\L.
\ee
We have (with $A$ as in (\ref{c2})): 
\be\label{c8}
\t{A}C^{\pm}A=diag\{\l_{1}^{\pm},\l_{2}^{\pm},...,\l_\L^{\pm}\}.
\ee
Corresponding to (\ref{c5}) we have
\be\label{c9}
(C^{\pm}A)_{ij}=\l_j^{\pm} A_{ij}.
\ee

Set $K^+\equiv J^+/kT$, with $J^+\equiv J$ (resp. $K^-\equiv J^-/kT$, 
with $J^-\equiv -J$ )
for ferromagnetic, i.e., with $J>0$ (resp. antiferromagnetic, i.e., 
with $J<0$) systems. From (\ref{cp}), (\ref{cm}) and
$\displaystyle\sum_{i=1}^\L\sigma_i^2=\L$, we see that the partition
function (\ref{2}) can be rewritten as
\be\label{9}
Z_\Lambda^{\pm}=exp\left\{{\pm}\l^{\pm} K^{\pm}\L\right\}
\displaystyle\sum_{\{\sigma_i\}}exp\left\{ -K^{\pm}
\displaystyle\sum_{i,j=1}^\Lambda C_{ij}^{\pm}\sigma_i\sigma_j
+K^\p\displaystyle\sum_{i=1}^\Lambda\sigma_i\right\},
\ee
where $Z_\Lambda^{+}$ (resp. $Z_\Lambda^{-}$) represents the partition
function for ferromagnetic (resp. antiferromagnetic) systems. In formula
(\ref{9}) all the interaction couplings could be included in the
elements of the matrix $C^\pm$. The parameter $J^\pm$ is no longer independent
and could be scaled to be $1$. $Z_\L^\pm$ is obviously independent of
$\epsilon$ by the definition of $\l^\pm$, (\ref{cp}) and (\ref{cm}).

Formula (\ref{9}) can be transformed into an integration by using the
following lemma, see e.g. \cite{gam}:

{\sf [Lemma].} Let $f$ and $S$ be real-valued measurable functions on 
$\Rb^d$ such that $F(\l)\equiv\int_{\Rb^d} f(x)exp\left\{\l
S(x)\right\}dx$ exists and $S\in C^2(\Rb^d)$.
For $0<\l\to\infty$, $F(\l)$ is equal to
\be\label{10}
F(\l)=(\displaystyle\frac{2\pi}{\l})^{d\over 2} \displaystyle\sum_{x^0}
\vert det S^\pp(x^0)
\vert^{-\frac{1}{2}}\left[f(x^0)+O(\displaystyle\frac{1}{\l})\right]exp\left
\{\l S(x^0)\right\},
\ee
provided $S(x)$ has a
finite number of non-degenerate maxima $x^0$
such that
$$
S^\p(x)\vert_{{x}^0}=\nabla S(x)\vert_{{x}^0}=0;
$$
$$
det S^\pp (x)\vert_{x^0}=
det\left.\left(\displaystyle\frac{\partial^2 S(x)}{\partial
x_i\partial x_j}\right)\right\vert_{{x}^0}\neq 0,~~~i,j=1,2,...,d
$$
and $max_{x\in \Rb^d} S(x)=S({x}^0)$. The sum in (\ref{10}) goes over all
such maxima $x^0$. \hfill $\rule{2mm}{2mm}$

Now let
\be\label{11}
S(x)=-\displaystyle\sum_{i=1}^d (x_i^2-1)^2,~~~~x_i\in\Rb,~~~~
x=(x_1,...x_d)\in\Rb^d.
\ee
Then the critical points $x^0$ of $S$ are given by 
\be\label{cri}
\left.\frac{\partial S(x)}{\partial x_i}\right\vert_{x^0}=-4(x_i^2-1)x_i
\vert_{{x}^0}=0,~~~~~i=1,...,d.
\ee
Here we have
\be\label{hess}
det S^\pp =det \left(\displaystyle\frac{\partial^2 S(x)}{\partial
x_i\partial x_j}\right)=det(-(12 x_i^2-4)\delta_{ij}).
\ee
From (\ref{cri}) and (\ref{hess}) we see that all critical points (maxima)
$x^0\in\Rb^d$ satisfying $det S^\pp\neq 0$ are given by 
$x^0=(x_i^0)$, $i=1,...,d$, for all possible combinations of values
$x^0_i=\pm 1$. We have $max_{x\in\Rb^d}S(x)=S(x^0)=0$.

Let ${\cal C}$ denote the set of all critical points.
By the {\sf Lemma} we get asymptotically for $\l\to\infty$,
$$
\int_{\Rb^d} f(x) exp\left\{-\l\displaystyle\sum_{i=1}^d (x_i^2-
1)^2\right\}dx=
(\displaystyle\frac{2\pi}{\l})^{d\over 2}8^{-\frac{d}{2}}
\left[\displaystyle\sum_{y\in {\cal C}}f(y)
+O(\displaystyle\frac{1}{\l})\right]
$$
for any $f$ such that the integration exists.

In particular we have
\be\label{parti}
\lim_{\l\to\infty}\l^{d\over 2}\int_{\Rb^d} f(x) 
exp\left\{-\l\displaystyle\sum_{i=1}^d (x_i^2-
1)^2\right\}dx=
(2\pi)^{d\over 2}8^{-\frac{d}{2}}
\displaystyle\sum_{y\in {\cal C}}f(y).
\ee
We shall now apply this result to the case $d=\L$, 
observing that $\sum_{y\in {\cal C}}f(y)$ amounts to sum over $y$ with
$y_i=\pm 1$ (independently, for each $i=1,...,\L$) 
and identifying then $y_i$ with the $\sigma_i$ occurring in (\ref{9}) we 
see that the partition function (\ref{9}) can be written as
\be\label{12}
\ba{rcl}
Z_\Lambda^{\pm}&=&\displaystyle
exp\left\{{\pm}\l^{\pm} K^{\pm}\L\right\}(\displaystyle\frac{\l}
{2\pi})^{\L\over 2}8^{\frac{\L}{2}}\cdot\\[5mm]
&&\displaystyle\int_{\Rb^\L}exp\left\{ -K^{\pm} \displaystyle
\sum_{i,j=1}^\Lambda C_{ij}^{\pm}y_i y_j
+K^\p\displaystyle\sum_{i=1}^\Lambda y_i-\l\displaystyle
\sum_{i=1}^\L(y_i^2-1)^2\right\}dy
\vert_{\l\to\infty},
\ea
\ee
with the notation $F(\l)\vert_{\l\to\infty}\equiv\lim_{\l\to\infty}F(\l)$.

We call (\ref{12}) the integral representation of the partition functions
$Z_\L^\pm$.
Now we shall study its properties. First however we make a remark:

{\sf Remark 1}. By combining the first two terms in the exponential of the
integrand in (\ref{12}) into a quadratic form and making
a translation of the integration variable $y_i\to y_i-a^\pm$ in
(\ref{12}), with
\be\label{a}
a^{\pm}\equiv\displaystyle\frac{-K^\p}{2K^{\pm}\sum_{i=1}^\L C_{ij}^{\pm}},
\ee
we get
$$
\ba{rcl}
Z_\Lambda^{\pm}
&=&exp\left\{{\pm}\l^{\pm} K^{\pm}\L\right\}
(\displaystyle\frac{\l}{2\pi})^{\L\over 2}8^{\frac{\L}{2}}
\displaystyle\int_{\Rb^\L}exp\left\{ -K^{\pm}
\displaystyle\sum_{i,j=1}^\Lambda C_{ij}^{\pm}(y_i+a^\pm)(y_j+a^\pm)
\right.\\[5mm]
&&\left.+K^{\pm}(a^{\pm})^2\displaystyle\sum_{i,j=1}^\Lambda C_{ij}^{\pm}
-\l\displaystyle\sum_{i=1}^\L(y_i^2-
1)^2\right\}dy\vert_{\l\to\infty}\\[5mm]
&=&exp\left\{{\pm}\l^{\pm} K^{\pm}\L\right\}
(\displaystyle\frac{\l}{2\pi})^{\L\over 2}8^{\frac{\L}{2}}
\displaystyle\int_{\Rb^\L}exp\left\{ -K^{\pm}
\displaystyle\sum_{i,j=1}^\Lambda C_{ij}^{\pm}y_iy_j\right.\\[5mm]
&&\left.+K^{\pm}(a^{\pm})^2\displaystyle\sum_{i,j=1}^\Lambda C_{ij}^{\pm}
-\l\displaystyle\sum_{i=1}^\L((y_i-a^{\pm})^2-
1)^2\right\}dy\vert_{\l\to\infty}.
\ea
$$
Therefore it is clear that the external magnetic field 
$H=K^\p kT=-2a^\pm J^\pm\displaystyle\sum_{i=1}^\L C_{ij}^{\pm}$ contributes a 
global translation $a^\pm$ of the critical points
of the function (\ref{11}) together with a constant factor 
$exp\left\{K^{\pm}(a^{\pm})^2\displaystyle
\sum_{i,j=1}^\Lambda C_{ij}^{\pm}\right\}$
to the zero-field partition functions $Z_\L^{\pm}(H=0)$:
\be\label{14}
\ba{rcl}
Z_\L^{\pm}(H=0)&=&
exp\left\{\pm\l^\pm K^\pm\L\right\}
(\displaystyle\frac{\l}{2\pi})^{\L\over 2}8^{\frac{\L}{2}}
\int_{\Rb^\L}exp\left\{ -K^\pm\displaystyle\sum_{i,j=1}^\Lambda C_{ij}^\pm 
y_i y_j\right.\\[5mm]
&&\left.-\l\displaystyle\sum_{i=1}^\L(y_i^2-1)^2\right\} dy\vert_{\l\to\infty}.
\ea
\ee

Conversely, a global translation of the critical points in 
the zero-field partition function is equivalent with
the introduction of an external magnetic field.

For convenience in Theorem 1 below we shall denote by $\L$ the number
of points in a one dimensional sublattice in an $n$-dimensional square lattice, 
so that $\L^n$ is the total number
of points of the given $n$-dimensional lattice. 
From (\ref{2}) it is obvious that for given $K$, $K^\p$ and lattice number $\L$,
the partition function $Z_\L$ is uniquely determined by the interaction coupling 
matrix $C^0$. Further we can prove the following:

{\sf [Theorem 1].} For given $K$, $K^\p$, 
the partition function $Z_{\L^n}$ of n-dimensional homogeneous Ising systems
(described by (\ref{2})) on a square lattice
with $\L^n$ lattice sites is uniquely given by the eigenvalues of the
corresponding interaction coupling matrix.

{\sf [Proof].} We first consider the one-dimensional homogeneous case. In
this case the interaction coupling matrix $C^0$ has the form
\be\label{c0}
C^0=\left(
\ba{cccc}
a_1&a_2&\cdots&a_\L\\[3mm]
a_\L&a_1&\cdots&a_{\L-1}\\[3mm]
\vdots&\vdots&\cdots&\vdots\\[3mm]
a_2&a_3&\cdots&a_1
\ea\right),
\ee
where $a_\alpha$, $\alpha=1,2,...,\L$ are real constants representing
the $(\alpha-1)$-th order nearest neighbours interactions. The diagonal
element $a_1$ is in fact zero, $a_1=0$. Since the system is homogeneous
we have
\be\label{aalpha}
a_\alpha=a_{\L-\alpha+2}\left\{
\ba{ll}
\alpha=2,3,...,\displaystyle\frac{\L}{2},~~~~\L~~even\\[3mm]
\alpha=2,3,...,\displaystyle\frac{\L+1}{2},~~~~\L~~odd
\ea\right.
\ee
This implies that $C^0$ is symmetric.

The matrix (\ref{c0})
is generated by cyclically permutating the elements of the former row to
the right. We call $C^0$ a right cyclic matrix. 
By using the $\L\times\L$ permutation matrix $P$,
\be\label{mp}
P=\left(
\ba{ccccc}
0&1&0&\cdots&0\\[1mm]
0&0&1&\cdots&0\\[1mm]
\vdots&\vdots&\vdots&\cdots&\vdots\\[1mm]
0&0&0&\cdots&1\\[1mm]
1&0&0&\cdots&0\\[1mm]
\ea\right),
\ee
the right cyclic matrix $C^0$ can be expressed as
\be\label{c0r}
C^0=a_1+a_2P+a_3P^2+\cdots+a_\L P^{\L-1}.
\ee

The $k$-th component of the $\alpha$-th eigenvector of the $\L\times\L$
permutation matrix $P$ acting in $\Cb^\L$ is simply given by
\be\label{ev}
exp\left\{\frac{2\pi i}{\L}(\alpha-1)(k-1)\right\},~~~~k,\alpha=1,2,...,\L,
\ee
see e.g., \cite{albeverio,grobner}.
As the eigenvalues of a real symmetric matrix are real, both the real
part and the imaginary part of the eigenvectors (\ref{ev}) are also
eigenvectors for the matrix $P$. We take
the $k$-th component of an eigenvector $u_\alpha$, $\alpha=1,2,...,\L$,
of $P$ to be normalized by the factor $1/\sqrt{\L}$ as follows,
\be\label{uo}
(u_\alpha)_k=\left\{
\ba{ll}
\displaystyle\frac{1}{\sqrt{\L}}\cos\left(\displaystyle\frac{2\pi}{\L}
(k-1)(\alpha -1)\right),
~~~~1\leq \alpha\leq \beta,\\[5mm]
\displaystyle\frac{1}{\sqrt{\L}}\sin\left(\displaystyle\frac{2\pi}{\L}
(k-1)(\alpha-1)\right),
~~~~\L\geq \alpha > \beta,
\ea\right.
\ee
where $k=1,2,...,\L$, $\beta=(\L+1)/2$ for $\L$ odd and 
$\beta=(\L+2)/2$ for $\L$ even. From (\ref{uo})
we see that $u_1$ is just the eigenvector (\ref{u}).

The orthogonal matrix that diagonalizes $P$ is then given by
\be\label{a1d}
A=(u_1,u_2,...,u_\L).
\ee
It is direct to check that the matrix $A$ obtained in this way satisfies 
the equations
(\ref{c4}) and (\ref{c10}). From (\ref{c0r}), (\ref{cm}) and
(\ref{cp}), the matrix $A$ also diagonalizes $C^\pm=\pm\l^\pm I\mp C^0$,
in the way given by formula (\ref{c8}).

Changing the integration variable $y_i$ to  be 
$\displaystyle\sum_{j=1}^\L A_{ij}y_j$ in the integration (\ref{12}) and using 
the relations (\ref{c4}) and (\ref{c8}) we get,
\be\label{t1z}
\ba{rcl}
Z_\Lambda^{\pm}&=&\displaystyle
exp\left\{{\pm}\l^{\pm} K^{\pm}\L\right\}
(\displaystyle\frac{\l}{2\pi})^{\L\over 2}
8^{\frac{\L}{2}}\displaystyle
\int_{\Rb^\L}exp\left\{ -K^{\pm} \displaystyle
\sum_{i=1}^\Lambda \l_{i}^{\pm}y_i^2\right.\\[5mm]
&&\left.+K^\p\displaystyle\sum_{ij=1}^\Lambda A_{ij}y_j
-\l\displaystyle\sum_{ijk=1}^\L(A_{ij}A_{ik}y_jy_k-1)^2\right\}
dy\vert_{\l\to\infty}.
\ea
\ee
From (\ref{uo}) and (\ref{a1d}), we see that $A=(A_{ij})$ is independent of the
elements of $C^\pm$. Therefore for
one dimensional homogeneous systems the partition function (\ref{t1z})
only depends the eigenvalues $\l_i^\pm$, i.e., the eigenvalues of the
interaction coupling matrix $C^0$.

For two dimensional homogeneous square lattice systems with $\L^2$ lattice 
sites, the interaction coupling matrix $C^0$, denoted here by $C_2^0$,
may have various forms according to the ways 
of numbering the lattice sites. We number the lattice sites
from left to right and from the first line to the last line of the lattice.
The matrix $C^0_2$ is then of the form,
\be\label{c02}
C^0_2=\left(
\ba{cccc}
C^0&a_2 I&\cdots&a_\L I\\[3mm]
a_\L I&C^0&\cdots&a_{\L-1} I\\[3mm]
\vdots&\vdots&\cdots&\vdots\\[3mm]
a_2 I&a_3 I&\cdots&C^0
\ea\right)=C^0\otimes I+I\otimes C^0,
\ee
where $C^0$ and $a_\alpha$, $\alpha=2,...,\L$, as in (\ref{c0}), 
and $I$ is the $\L\times\L$ identity matrix.

$C^0_2$ is again a right cyclic matrix generated by permuting the matrix blocks
($C^0$,$a_2 I$, $a_3 I$,...,$a_\L I$). The eigenvectors of the interaction coupling
matrix $C_2^0$ have the form,
\be\label{v}
v_{\alpha\beta}=u_\alpha\otimes u_\beta,~~~~~\alpha,\beta=1,2,...,\L,
\ee
with $u_\alpha$, $u_\beta$ as in (\ref{uo}). The matrix that diagonalizes
$C^0_2$ is then given by
\be\label{va}
A=(v_{11},v_{12},...,v_{21},v_{22},...,v_{\L\L}).
\ee
$A$ is then a $\L^2\times\L^2$ othonormal matrix which is independent of 
$a_2,...,a_\L$. Obviously the matrix ({\ref{va}})
also diagonalizes the positive definite matrix $C_2^\pm=\pm\l^\pm I\otimes 
I\mp C^0_2$ in the integral representation of the partition
functions. Therefore for two-dimensional homogeneous systems the
partition functions are also only determined by the eigenvalues of the
interaction coupling matrix for given $K$, $K^\p$ and $\L$.

Similarly, for an $n(>2)$-dimensional homogeneous system, the interaction
coupling matrix $C^0_n$ has the form,
\be\label{c0n}
C^0_n=\left(
\ba{cccc}
C^0_{n-1}&a_2 I&\cdots&a_\L I\\[3mm]
a_n I&C^0_{n-1}&\cdots&a_{\L-1} I\\[3mm]
\vdots&\vdots&\cdots&\vdots\\[3mm]
a_2 I&a_3 I&\cdots&C^0_{n-1}
\ea\right)=C^0_{n-1}\otimes I+I\otimes C^0_{n-1},
\ee
where $C_{n-1}^0$ is the interaction coupling matrix of an $(n-1)$-dimensional
homogeneous system. 

The matrix that diagonalizes $C^0_n$ and $C_n^\pm=\pm\l^\pm I_n\mp
C^0_n$, with $I_n$ the $\L^n\times\L^n$ identity matrix, is given by
\be\label{a0n}
A=(u_{\alpha_1}\otimes u_{\alpha_2}\otimes ...\otimes
u_{\alpha_n}),~~~~\alpha_i=1,2,...,\L,~~~~\forall i=1,2,...,n,
\ee
where $u_{\alpha_i}$ is given by (\ref{uo}). This proves the theorem.
\hfill $\rule{2mm}{2mm}$

{\sf [Theorem 2].} For one-dimensional homogeneous ferromagnetic Ising systems
with interaction coupling matrix $C^0$ 
given by (\ref{c0}) and coupling coefficients $a_i>0$, $i=2,3,...,\L$, 
the partition function $Z_\L^+$ satisfies
\be\label{t2}
\sum_{i=2}^{\L}a_i\displaystyle\frac{\partial Z_\L^+}{\partial a_i}
+T\displaystyle\frac{\partial Z_\L^+}{\partial T}
+H\displaystyle\frac{\partial Z_\L^+}{\partial H}
+L\sum_{i=2}^{\L}\displaystyle\frac{\partial Z_\L^+}{\partial a_i}
=K^+\L (\L-1) L Z_{\L}^+.
\ee

{\sf [Proof].} From (\ref{12}) the partition function of one-dimensional
ferromagnetic homogeneous systems takes the form,
\be\label{tnz}
\ba{rcl}
Z_\L^+&=&\displaystyle
exp\left\{\l^{+} K^{+}\L\right\}(\displaystyle\frac{\l}{2\pi})^{{\L}\over 2}
8^{\frac{{\L}}{2}}\displaystyle
\int_{\Rb^{\L}}exp\left\{ 
-K^{+} \displaystyle\sum_{i=1}^{\L} \l_{i}^{+}y_i^2\right. \\[5mm]
&&\left. +K^\p\displaystyle\sum_{ij=1}^{\L} A_{ij}y_j
-\l\displaystyle\sum_{ijk=1}^{\L}(A_{ij}A_{ik}y_jy_k-1)^2\right\}
dy\vert_{\l\to\infty}\\[5mm]
&\equiv&\displaystyle
B^+\int_{\Rb^{\L}}exp\left\{\Gamma^+\right\}
dy\vert_{\l\to\infty},
\ea
\ee
where $A_{ij}$, $i,j=1,2,...,{\L}$ are the elements of the matrix
(\ref{a1d}), $\l_i^{+}=\l^{+}-\l^0_i=\l^+_{max}+\epsilon-\l^0_i$,
$\l^0_i$ is an eigenvalue of $C^0$. For simplicity we have set
in (\ref{tnz}),
\be\label{bppp}
B^+\equiv exp\left\{\l^{+} K^{+}\L\right\}
(\displaystyle\frac{\l}{2\pi})^{{\L}\over 2} 8^{\frac{{\L}}{2}},
\ee
\be\label{gppp}
\Gamma^+\equiv
-K^{+} \displaystyle\sum_{i=1}^{\L} \l_{i}^{+}y_i^2
+K^\p\displaystyle\sum_{ij=1}^{\L} A_{ij}y_j
-\l\displaystyle\sum_{ijk=1}^{\L}(A_{ij}A_{ik}y_jy_k-1)^2.
\ee

By (\ref{9}) and the
definitions (\ref{cp}), (\ref{cm}) of $C^+$, as we stated before, the partition
function $Z_\L^+$ is independent of the positive real number $\epsilon$.
Hence we have
\be\label{ppp}
0=\displaystyle\frac{\partial Z_\L^+}{\partial\epsilon}.
\ee
On the other hand using Lebsgue dominated convergence one shows that the
application of $\partial/\partial\epsilon$ to the right hand side of
(\ref{tnz}) yields
\be\label{pepsi}
\ba{l}
K^+\L Z_\L^++\displaystyle
exp\left\{\l^{+} K^{+}\L\right\}(\displaystyle\frac{\l}{2\pi})^{{\L}\over 2}
8^{\frac{{\L}}{2}}\cdot\\[5mm]
\displaystyle
\int_{\Rb^{\L}}exp\left\{ -K^{+} \displaystyle\sum_{i=1}^{\L} \l_{i}^{+}y_i^2 
+K^\p\displaystyle\sum_{ij=1}^{\L} A_{ij}y_j
-\l\displaystyle\sum_{ijk=1}^{\L}(A_{ij}A_{ik}y_jy_k-1)^2\right\}\\[5mm]
\cdot\left[- K^{+} \displaystyle\sum_{i=1}^{\L}y_i^2  \right]
dy\vert_{\l\to\infty},
\ea
\ee
(In fact the derivatives of $Z_\L^+$ with respective to $\epsilon$ and, in
the following, to $T$, $H$ and $a_i$, can be exchanged with the integration
and the limitation in (\ref{tnz}) because from (\ref{9}) these derivatives only change
the form of the function $f$ in (\ref{parti})).
Equating (\ref{pepsi}) with (\ref{ppp}) we get
\be\label{pe}
\L Z_\L^+ - B^+
\int_{\Rb^{\L}}exp\left\{\Gamma^+\right\}
\displaystyle\sum_{i=1}^{\L}y_i^2
dy\vert_{\l\to\infty}=0.
\ee

In a similar way we get from (\ref{tnz})
\be\label{ph}
\displaystyle\frac{\partial Z_\L^+}{\partial H}=B^+
\int_{\Rb^{\L}}exp\left\{\Gamma^+\right\}
\left[\displaystyle\frac{1}{kT} \displaystyle\sum_{ij=1}^{\L}A_{ij}y_j  \right]
dy\vert_{\l\to\infty}.
\ee
and
\be\label{pt}
\ba{rcl}
\displaystyle\frac{\partial Z_\L^+}{\partial T}&=&
\l^+ \L \left(-\displaystyle\frac{J^+}{k T^2}\right)Z_\L^+\\[5mm]
&&+B^+\displaystyle\int_{\Rb^{\L}}exp\left\{\Gamma^+\right\}
\left[\displaystyle\frac{J^+}{kT^2}
\displaystyle\sum_{i=1}^{\L}\l_{i}^+ y_i^2
-\displaystyle\frac{H}{kT^2}
\displaystyle\sum_{ij=1}^{\L}A_{ij}y_j  \right]
dy\vert_{\l\to\infty}.
\ea
\ee
From (\ref{ph}) and (\ref{pt}) we deduce that
\be\label{pth}
\ba{rcl}
\displaystyle\frac{\partial Z_\L^+}{\partial T}
+\displaystyle\frac{H}{T}\displaystyle\frac{\partial Z_\L^+}{\partial H}
&=&\l^+ \L \left(-\displaystyle\frac{K^+}{T}\right)Z_\L^+\\[5mm]
&&+B^+\displaystyle\int_{\Rb^{\L}}exp\left\{\Gamma^+\right\}
\left[\displaystyle\frac{K^+}{T}
\displaystyle\sum_{i=1}^{\L}\l_{i}^+ y_i^2\right]
dy\vert_{\l\to\infty}.
\ea
\ee
Using (\ref{pe}) one gets
\be\label{pthe}
T\displaystyle\frac{\partial Z_\L^+}{\partial T}+
H\displaystyle\frac{\partial Z_\L^+}{\partial H}=
B^+\int_{\Rb^{\L}}exp\left\{\Gamma^+\right\}
\left[- K^+
\displaystyle\sum_{i=1}^{\L}\l^0_i y_i^2\right]
dy\vert_{\l\to\infty}.
\ee

Now we discuss the properties of the eigenvalues of the matrix $C^0$ 
given by (\ref{c0}). We first recall a general result of matrix theory.
Let $g(x)$ be polynomial in $x$ of degree $l$ with roots
($\rho_1,\rho_2,...,\rho_l$),
\be\label{gx}
\ba{rcl}
g(x)&=&d_0 x^l+d_1 x^{l-1}+...+d_{l-1} x+d_l\\[4mm]
&=&(-1)^l d_0 (\rho_1-x)(\rho_2-x)...(\rho_l-x).
\ea
\ee
Let $\l_i$, $i=1,2,...,m$, be the eigenvalues of a given $m\times m$ matrix
$B$ over $\Cb$, acting in $\Cb^m$. Then
$$
\vert\rho I-B\vert=(\rho-\l_1)(\rho-\l_2)...(\rho-\l_m).
$$
Replacing $x$ by $B$ in (\ref{gx}) we get
$$
g(B)=(-1)^l d_0 (\rho_1 I-B)(\rho_2 I-B)...(\rho_l I-B).
$$
Hence
\be\label{gbd}
\ba{rcl}
\vert g(B)\vert &=&(-1)^{ml} d_0^m \vert (\rho_1 I-B)\vert 
\vert (\rho_2 I-B)\vert ...\vert (\rho_l I-B)\vert\\[4mm]
&=&(-1)^{ml} d_0^m \displaystyle\prod_{k=1}^{l}\displaystyle
\prod_{i=1}^{m}(\rho_k-\l_i)=\displaystyle\prod_{i=1}^{m}g(\l_i),
\ea
\ee
see e.g. \cite{grobner}.

From (\ref{c0r}) we see that $C^0$ is a polynomial in the $\L\times\L$
permutation matrix $P$ acting in $\Cb^\L$. 
As $\vert \rho I- P\vert=\rho^\L-1$, the eigenvalues of $P$ are
\be\label{zeta}
\zeta_\alpha=exp\left\{\frac{2\pi i(\alpha-1)}{\L}\right\},
~~~~\alpha=1,2,...,\L.
\ee
Noting that all diagonal elements of $C^0$ are equal to $a_1$, 
from (\ref{gbd}) we have
\be\label{dc0}
\ba{rcl}
\vert C^0-\l^0 I\vert &=&\vert a_1-\l^0 + a_2 P+ a_3 P^2+...+a_\L P^{\L-
1}\vert\\[4mm]
&=&\displaystyle\prod_{\alpha=1}^\L\left[a_1-\l^0+ 
a_2\zeta^\alpha+ a_3\zeta^{2\alpha}+...+ a_\L\zeta^{(\L-1)\alpha}\right].
\ea
\ee
The eigenvalues of $C^0$ are then given by
\be\label{lc0}
\l_\alpha^0=a_2\zeta^\alpha+ a_3\zeta^{2\alpha}+...
+ a_\L\zeta^{(\L-1)\alpha},~~~~\alpha=1,2,...,\L.
\ee

By the definition (\ref{link}) we see that $\l_1^0=
\sum_{\alpha=2}^\L a_\alpha$ is
of special significance. It is just the generalized link number of the
lattice, $L=\l_1^0$. For our case, $a_\alpha>0$, $\alpha=2,3,...,\L$, and 
the link number $L$ is also the largest eigenvalue of the interaction 
coupling matrix $C^0$,
\be\label{lmax}
\l^+_{max}=\l^0_1=L,
\ee
which means $\l^+_1=\l^+_{max}+\epsilon -\l_1^0=\epsilon$ and hence
\be\label{hhh}
\frac{\partial \l^+_1}{\partial \l_i^0}=0,~~~~i=1,2,...,\L.
\ee

From (\ref{lc0}) $\l_\alpha^0$ satisfies
\be\label{l1}
\sum_{\gamma=2}^\L a_\gamma\displaystyle\frac{\partial \l_\alpha^0}{\partial
a_\gamma}=\l_\alpha^0,
\ee
\be\label{l2}
\Theta_\alpha\equiv
\sum_{\gamma=2}^\L \displaystyle\frac{\partial \l_\alpha^0}{\partial a_\gamma}=
\left\{\ba{l}\L-1,~~~~\alpha=1,\\[3mm]-1,~~~~~~
\alpha\neq 1.\ea\right.
\ee
The eigenvector corresponding to $\l_\alpha^0$ is $u_\alpha$ given by
(\ref{uo}),
\be\label{eig}
C^0 u_\alpha=\l_\alpha ^0 u_\alpha.
\ee

From (\ref{tnz}), (\ref{l1}) and (\ref{hhh}) we have,
\be\label{apa}
\ba{rcl}
\displaystyle\sum_{i=2}^n a_i \displaystyle\frac{\partial Z_\L^+}{\partial a_i}
&=&\displaystyle\sum_{i=2}^n\displaystyle\sum_{\mu=1}^{\L}
a_i \displaystyle\frac{\partial Z_\L^+}{\partial \l^0_\mu}
\displaystyle\frac{\partial \l^0_\mu}{\partial a_i}
=\sum_{\mu=1}^{\L}
\displaystyle\frac{\partial Z_\L^+}{\partial \l^0_\mu}\l^0_\mu\\[5mm]
&=&L K^+ \L Z_\L^+
+B^+\displaystyle\int_{\Rb^{\L}}exp\left\{\Gamma^+\right\}
\left[K^+ \displaystyle\sum_{i=2}^{\L}\l^0_{i} y_i^2\right]
dy\vert_{\l\to\infty},
\ea
\ee
where the relation (\ref{lmax}) has been used.

By (\ref{tnz}), (\ref{l2}) and (\ref{hhh}) we also have
\be\label{pa}
\ba{rcl}
\displaystyle\sum_{i=2}^n \displaystyle\frac{\partial Z_\L^+}{\partial a_i}
&=&\displaystyle\sum_{i=2}^n\sum_{\mu=1}^{\L}
\displaystyle\frac{\partial Z_\L^+}{\partial \l^0_\mu}
\displaystyle\frac{\partial \l^0_\mu}{\partial a_i}
=\sum_{\mu=1}^{\L}
\displaystyle\frac{\partial Z_\L^+}{\partial \l^0_\mu}
\Theta_\mu\\[5mm]
&=&K^+ \L (\L-1) Z_\L^+
-B^+\displaystyle\int_{\Rb^{\L}}exp\left\{\Gamma^+\right\}
\left[K^+ \displaystyle\sum_{i=2}^{\L} y_i^2\right]
dy\vert_{\l\to\infty}.
\ea
\ee

From (\ref{pthe}), (\ref{apa}) one has
\be\label{thaa}
\ba{rcl}
\displaystyle\sum_{i=1}^{\L}a_i \displaystyle
\frac{\partial Z_\L^+}{\partial a_i}+
T\displaystyle\frac{\partial Z_\L^+}{\partial T}+
H\displaystyle\frac{\partial Z_\L^+}{\partial H}&=&
L K^+\L Z_\L^+\\[5mm]
&&+B^+\displaystyle\int_{\Rb^{\L}}exp\left\{\Gamma^+\right\}
(- K^+ L y_1^2)dy\vert_{\l\to\infty}.
\ea
\ee
By (\ref{pe}) and (\ref{pa}) one gets
\be\label{pae}
\sum_{i=2}^{\L}\displaystyle\frac{\partial Z_\L^+}{\partial a_i}
=K^+\L(\L-2) Z_\L^+
+ K^+ B^+\int_{\Rb^{\L}}exp\left\{\Gamma^+\right\}y_1^2
dy\vert_{\l\to\infty}.
\ee
Equations (\ref{thaa}) and (\ref{pae}) then give
$$
\ba{l}
\displaystyle\sum_{i=2}^{\L}a_i \displaystyle
\frac{\partial Z_\L^+}{\partial a_i}+
T\displaystyle\frac{\partial Z_\L^+}{\partial T}+
H\displaystyle\frac{\partial Z_\L^+}{\partial H}
+L\displaystyle\sum_{i=2}^{\L}\displaystyle\frac{\partial
Z_\L^+}{\partial a_i}\\[5mm]
=L K^+\L Z_\L^+
+L K^+\L(\L-2) Z_\L^+=L K^+\L Z_\L^+(\L-1),
\ea
$$
which is just the equality (\ref{t2}). \hfill $\rule{2mm}{2mm}$

{\sf Remark 2}. Equality (\ref{t2}) means that the variation of the 
interaction
couplings of a system is related to the variations of the external
magnetic field and the temperature. As the free energy $F^+$, internal 
energy $u^+$ and magnetization $M^+$ per site are given respectively by
\be\label{fp}
F^+=-\frac{1}{\L}kT\log Z^+,
\ee
\be\label{up}
u^+=-T^2\frac{\partial}{\partial T}\left(-\frac{k}{\L}\log Z^+\right)
=\frac{T^2 k}{\L}\frac{\partial}{\partial T}\log Z^+,
\ee
\be\label{map}
M^+=-\frac{\partial F^+}{\partial H}
=\frac{T k}{\L}\frac{\partial}{\partial H}\log Z^+,
\ee
(\ref{t2}) can be rewritten as
\be\label{t2t2}
u^+ + M^+ H -\sum_{i=2}^{\L}(a_i+L)\frac{\partial F^+}{\partial a_i}
=J^+(\L-1)L.
\ee
(\ref{t2t2}) gives the relation among the internal energy,
magnetization, external field, free energy and interaction couplings
of one-dimensional homogeneous Ising ferromagnetic systems.

{\sf [Theorem 3].} For one-dimensional
homogeneous Ising ferromagnetic systems with interaction coupling matrix $C^0$
given by (\ref{c0}) and coupling coefficients $a_i>0$, $i=2,3,...,\L$,
the correlation function $g_{ij}$ between spins $i$ and $j$ satisfies
\be\label{at3}
\sum_{i=2}^{\L}a_i\displaystyle\frac{\partial g_{ij}}{\partial a_i}
+T\displaystyle\frac{\partial g_{ij}}{\partial T}
+H\displaystyle\frac{\partial g_{ij}}{\partial H}
+L\sum_{i=2}^{\L}\displaystyle\frac{\partial g_{ij}}{\partial a_i}=0.
\ee

{\sf [Proof].} Let $f(\sigma)$ be an entire function of $\sigma_i$,
$i=1,2,...,\L$, independent of $a_i$, $T$ and $H$. The average $<f(\sigma)>$ 
of $f(\sigma)$ is by definition given by 
\be\label{t3p}
<f(\sigma)>\equiv\displaystyle\frac{1}{Z_\L}
\displaystyle\sum_{\{\sigma_i\}}f(\sigma)exp\left\{K
\displaystyle\sum_{i,j=1}^\Lambda C^{0}_{ij}\sigma_i\sigma_j
+K^\p\displaystyle\sum_{i=1}^\Lambda\sigma_i\right\},
\ee
where $Z_\L$ is given by (\ref{2}).

For ferromagnetic systems, using our integral
representation we see that we can write (\ref{t3p}) in the form: 
\be\label{t3p1}
<f(\sigma)>
=\displaystyle\frac{1}{Z_\L^+}
B^+\displaystyle\int_{\Rb^{\L}}f(\sigma)exp\left\{\Gamma^+\right\}
dy\vert_{\l\to\infty}\equiv \displaystyle\frac{Z_\L^+(f)}{Z_\L^+},
\ee
where $Z_\L^+$, $B^+$ and $\Gamma^+$ are given by (\ref{tnz}),
(\ref{bppp}) and (\ref{gppp}) respectively, and
\be\label{zf}
Z_\L^+(f)\equiv
B^+\displaystyle\int_{\Rb^{\L}}f(\sigma)exp\left\{\Gamma^+\right\}
dy\vert_{\l\to\infty}.
\ee

Let $U$ be a linear operator defined on smooth functions of $a_i$,
$T$ and $H$ by
\be\label{lop}
U=\sum_{i=2}^{\L}a_i\displaystyle\frac{\partial}{\partial a_i}
+T\displaystyle\frac{\partial}{\partial T}
+H\displaystyle\frac{\partial}{\partial H}
+L\sum_{i=2}^{\L}\displaystyle\frac{\partial}{\partial a_i}.
\ee
From Theorem 2 we see that
\be\label{oz}
U Z_\L^+=K^+\L (\L-1) L Z_{\L}^+.
\ee
It is also straightforward to check that
\be\label{ozf}
U Z_\L^+(f)=K^+\L (\L-1) L Z_{\L}^+(f).
\ee

By (\ref{t3p1}), (\ref{oz}) and (\ref{ozf}) we get
\be\label{of}
\ba{rcl}
U<f(\sigma)>&=&\displaystyle\frac{1}{Z_\L^+}U Z_\L^+(f)
-\displaystyle\frac{Z_\L^+(f)}{(Z_\L^+)^2}U Z_\L^+\\[4mm]
&=&\displaystyle\frac{1}{Z_\L^+}K^+\L (\L-1) L Z_{\L}^+(f)
-\displaystyle\frac{Z_\L^+(f)}{(Z_\L^+)^2}K^+\L (\L-1) L
Z_{\L}^+=0.
\ea
\ee
For the correlation function
$g_{ij}\equiv <\sigma_i\sigma_j>-<\sigma_i><\sigma_j>$, we get using
the result (\ref{of}):
$$
Ug_{ij}=U<\sigma_i\sigma_j>-U(<\sigma_i>)<\sigma_j>-
<\sigma_i>U(<\sigma_j>)=0,
$$
which is just the formula (\ref{at3}). \hfill $\rule{2mm}{2mm}$

By using our integral approach to the Ising model, we have discussed some
properties of the partition function for a one-dimensional homogeneous case.
We now remark that
for some zero magnetic field cases, in arbitrary dimensions, 
the partition function can be 
given exactly in terms of special functions.

Let $A$ be the matrix that diagonalizes $C^\pm$ as in (\ref{c8}). Let us
consider the Ising system described by (\ref{9}).

{\sf [Theorem 4].} If $C^\pm$ is such that $A$ satisfies
\be\label{38}
\displaystyle\sum_{i=1}^\L
A_{ij}A_{ik}A_{im}A_{in}=\delta_{jk}\delta_{km}\delta_{mn}b_n
\ee
for some constants $b_n$, $Re b_n>0$, $n=1,2,...,\L$, then
\be\label{39}
\ba{rcl}
Z_\L^\pm(H=0)&=&exp\left\{(\pm\l^\pm K^\pm-\l)\L\right\}(\displaystyle
\displaystyle\frac{\l}{\pi})^{\L\over 2}
exp\left\{\displaystyle\sum_{i=1}^\L\displaystyle
\displaystyle\frac{(K^\pm)^2(\l_i^\pm-2\l)^2}{8\l b_i}\right\}\\[5mm]
&&\displaystyle\prod_{i=1}^\L\sqrt{\displaystyle\frac{K(\l_i^\pm-2\l)}{\l b_i}}
K_{\frac{1}{4}}\left({\displaystyle\frac{K^2(\l_i^\pm-2\l)^2}{8\l b_i}}\right)
\vert_{\l\to +\infty},
\ea
\ee
where $K_{\frac{1}{4}}$ is the Bessel function of imaginary argument equal to 
$\frac{1}{4}$.

{\sf [Proof].} The partition function of zero-field is given by
(\ref{14}). Set $y_i=\displaystyle\sum_{j=1}^\Lambda A_{ij}p_j$ in (\ref{14}),
$A=(A_{ij})$ as in (\ref{c2}). By using (\ref{c8}), (\ref{c4}) and
(\ref{38}) we have
\be\label{40}
\displaystyle\sum_{i,j=1}^\Lambda C_{ij}^\pm y_iy_j=\displaystyle\sum_{i=1}^\Lambda 
\l_i^\pm p_i^2,
\ee
\be\label{41}
\displaystyle\sum_{i=1}^\L y_i^2=\displaystyle\sum_{i=1}^\L 
(\displaystyle\sum_{j=1}^\Lambda A_{ij}p_j)^2
=\displaystyle\sum_{j,k=1}^\Lambda \displaystyle\sum_{i=1}^\L  
A_{ij}A_{ik}p_j p_k
=\displaystyle\sum_{i=1}^\L p_i^2
\ee
and
\be\label{42}
\displaystyle\sum_{i=1}^\L y_i^4=\displaystyle\sum_{i=1}^\L 
\displaystyle\sum_{j,k,m,n=1}^\Lambda
A_{ij}A_{ik}A_{im}A_{in}p_jp_kp_mp_n=\displaystyle\sum_{i=1}^\L b_i p_i^4.
\ee
Substituting $y_i=\displaystyle\sum_{j=1}^\Lambda A_{ij}p_j$,
(\ref{40}), (\ref{41}) and (\ref{42}) into (\ref{14}) we get
$$
\ba{rcl}
Z_\L^\pm(H=0)&=&
exp\left\{(\pm\l^\pm K^\pm-\l)\L\right\}(\displaystyle\frac{\l}{2\pi})^{\L\over 2}
8^{\frac{\L}{2}}\\[5mm]
&&\int_{\Rb^\L}exp\left\{ -K^\pm \displaystyle
\sum_{i=1}^\Lambda (\l_i^\pm-2\l) p_i^2
-\l\displaystyle\sum_{i=1}^\L b_i p_i^4\right\}dp\vert_{\l\to\infty}.
\ea
$$

From the formula (see e.g. \cite{table})
\be\label{43}
\int_{\Rb}exp\left\{ -\mu x^4-2\nu x^2\right\}dx=
\displaystyle\frac{1}{2}
\sqrt{\displaystyle\frac{2\nu}{\mu}}
exp\left\{\displaystyle\frac{\nu^2}{2\mu}\right\}K_{\frac{1}{4}}
\left(\displaystyle\frac{\nu^2}{2\mu}\right),~~~~Re\mu>0,
\ee
we obtain (\ref{39}). \hfill $\rule{2mm}{2mm}$

In the following we present an asymptotic formula for the partition
function (\ref{2}) in the general case. We first make a Laplace
transformation to the integral representation of the partition
function (\ref{12}). 

From the formula
\be\label{15}
\int_{\Rb^\L}exp\left\{ -K^\pm 
\displaystyle\sum_{i,j=1}^\Lambda B_{ij} p_i p_j\right\} dp
=\left({\pi\over K^\pm}\right)^{\L\over 2}\displaystyle\frac{1}{\sqrt{det B}}
\ee
for any symmetric strictly positive definite matrix $B$, 
(\ref{12}) can be reexpressed as
$$
\ba{rcl}
Z_\L^\pm&=&c^\pm \l^{\L\over 2}
\displaystyle\int_{\Rb^{\L}}\int_{\Rb^{\L}}
exp\left\{ -K^\pm \displaystyle\sum_{i,j=1}^\Lambda C_{ij}^\pm p_ip_j
+2iK^\pm\displaystyle\sum_{i,j=1}^\Lambda 
C_{ij}^\pm p_i (y_j+a^\pm)\right.\\[5mm]
&&\left.-\l\displaystyle\sum_{i=1}^\L(y_i^2-1)^2\right\}dpdy\vert_{\l\to\infty},
\ea
$$
where $a^\pm$ as in (\ref{a}) and
\be\label{16}
c^\pm\equiv exp\left\{\pm\l^\pm K^\pm\L\right\}
(\displaystyle\frac{1}{2\pi})^{\L\over 2}8^{\frac{\L}{2}}
exp\left\{K^\pm (a^\pm)^2\displaystyle\sum_{i,j=1}^\Lambda C_{ij}^\pm\right\}
({K^\pm\over \pi})^{\L\over 2}\sqrt{det C^\pm}.
\ee

By a translation $y_i\to y_i+1$ and a rescaling $y_i\to y_i/\sqrt{\l}$ we get
\be\label{17}
\ba{rcl}
Z_\Lambda^\pm
&=&c^\pm \l^{\L\over 2}\displaystyle\int_{\Rb^{\L}}\int_{\Rb^{\L}}
exp\left\{ -K^\pm \displaystyle\sum_{i,j=1}^\Lambda C_{ij}^\pm p_ip_j
+2iK^\pm\displaystyle\sum_{i,j=1}^\Lambda (a^\pm+1)C_{ij}^\pm p_i\right.\\[5mm] 
&&\left.+2iK^\pm\displaystyle\sum_{i,j=1}^\Lambda C_{ij}^\pm p_i y_j
-\l\displaystyle\sum_{i=1}^\L(y_i^4+4y_i^3+4y_i^2)\right\}
dpdy\vert_{\l\to\infty}\\[5mm]
&=&c^\pm\displaystyle\int_{\Rb^{\L}}\int_{\Rb^{\L}}
exp\left\{ -K^\pm \displaystyle\sum_{i,j=1}^\Lambda C_{ij}^\pm p_ip_j
+2iK^\pm\displaystyle\sum_{i,j=1}^\Lambda (a^\pm+1)C_{ij}^\pm p_i \right.\\[5mm] 
&&\left.+\displaystyle\frac{2iK^\pm}{\sqrt{\l}}
\displaystyle\sum_{i,j=1}^\Lambda C_{ij}^\pm p_i y_j
-\displaystyle\sum_{i=1}^\L(\displaystyle
\frac{y_i^4}{\l}+\displaystyle\frac{4y_i^3}{\sqrt{\l}}
+4y_i^2)\right\}dpdy\vert_{\l\to\infty}.
\ea
\ee

{\sf [Theorem 5].} For given $K$, $K^\p$, up to order $O(1/\l^2)$, 
the partition function of any $n$-dimensional homogeneous Ising
system is given in terms of the link number $L$ and the largest positive
eigenvalue $\l^+$ (resp. the largest absolute value $\l^-$ of the negative
eigenvalues) of the related interaction coupling matrix for ferromagnetic
(resp. antiferromagnetic) systems, according to the formula
\be\label{t4}
\ba{rcl}
Z_\L^\pm &=&exp\left\{(\pm K^\pm L+K^\p)\L\right\}
\left[1+\displaystyle\frac{\L}{\l}\left(
\displaystyle\frac{3}{16}+\displaystyle\frac{3}{16}
(2K^\pm(\pm\l^\pm \mp L)-K^\p)\right.\right.\\[5mm]
&&\left.\left.+\displaystyle\frac{1}{16}(2K^\pm(\pm\l^\pm\mp L)-K^\p)^2
\mp\displaystyle\frac{K^\pm}{8}\l^\pm\right)
\right]+O(\displaystyle\frac{1}{\l^2}).
\ea
\ee

{\sf [Proof].} By expanding the integrand in (\ref{17}) to order 
$\displaystyle\frac{1}{\l}$, before taking the limit $\l\to\infty$,
and using the integration formulae
\be\label{19}
\int_{-\infty}^{\infty} x^{2n}exp\{-px^2\}dx=\displaystyle\frac{(2n-
1)!!}{(2p)^n}\sqrt{\displaystyle\frac{\pi}{p}},~~~~p>0,
\ee
where $(2n-1)!!=1\cdot 3\cdot 5 \cdot...\cdot (2n-1)$, and
\be\label{20}
\int_{-\infty}^{\infty} x^{2n+1}exp\{-px^2\}dx=0,~~~~p>0,
\ee
we have
$$
\ba{rcl}
Z_\Lambda^\pm
&=&c^\pm\displaystyle\int_{\Rb^{\L}}\int_{\Rb^{\L}}exp\left\{ -K^\pm 
\displaystyle\sum_{i,j=1}^\Lambda C_{ij}^\pm p_ip_j
+2iK^\pm\displaystyle\sum_{i,j=1}^\Lambda (a^\pm+1)C_{ij}^\pm p_i \right\}\\[5mm] 
&&exp\left\{-4y_i^2)\right\}\left[
1+\displaystyle\frac{2iK^\pm}{\sqrt{\l}}\displaystyle
\sum_{i,j=1}^\Lambda C_{ij}^\pm p_i y_j-
\displaystyle\sum_{i=1}^\Lambda\left(\displaystyle
\displaystyle\frac{y_i^4}{\l}+\displaystyle\frac{4y_i^3}{\sqrt{\l}}\right)
\right.\\[5mm]
&&\left.+\displaystyle\frac{1}{2!\l}\left(-4(K^{\pm})^2\displaystyle
\displaystyle\sum_{i,j,k,l=1}^\Lambda C_{ij}^\pm C_{kl}^\pm p_i p_k y_jy_l
+16\displaystyle\sum_{i,j=1}^\Lambda y_i^3y_j^3\right.\right.\\[5mm]
&&\left.\left.-16iK^\pm\displaystyle\sum_{i,j,k=1}^\Lambda C_{ij}^\pm p_i
y_j y_k^3\right)\right]dpdy+O(\displaystyle\frac{1}{\l^2})\\[5mm]
&=&c^\pm\displaystyle\int_{\Rb^{\L}}\int_{\Rb^{\L}}exp\left\{ -K^\pm 
\displaystyle\sum_{i,j=1}^\Lambda C_{ij}^\pm p_ip_j
+2iK^\pm\displaystyle\sum_{i,j=1}^\Lambda (a^\pm+1)C_{ij}^\pm p_i \right\} 
exp\left\{-4y_i^2)\right\}\\[5mm]
&&\left[1-\displaystyle\sum_{i=1}^\Lambda\displaystyle
\displaystyle\frac{y_i^4}{\l}
+\displaystyle\frac{1}{2\l}\left(-4(K^{\pm})^2\displaystyle
\displaystyle\sum_{i,j,k=1}^\Lambda C_{ij}^\pm 
C_{kj}^\pm p_i p_k y_j^2\right.\right.\\[5mm]
&&\left.\left.+16\displaystyle\sum_{i=1}^\Lambda y_i^6
-16iK^\pm\displaystyle\sum_{i,j=1}^\Lambda C_{ij}^\pm p_i
y_j^4\right)\right]dpdy+O(\displaystyle\frac{1}{\l^2})\\[5mm]
&=&c^\pm\displaystyle\int_{\Rb^{\L}}exp\left\{ -K^\pm 
\displaystyle\sum_{i,j=1}^\Lambda C_{ij}^\pm p_ip_j
+2iK^\pm\displaystyle\sum_{i,j=1}^\Lambda (a^\pm+1)C_{ij}^\pm p_i
\right\}\\[5mm]
&&\left(\displaystyle\frac{\sqrt{\pi}}{2}\right)^\L
\left[1+\displaystyle\frac{1}{\l}\left(
\displaystyle\frac{3}{16}\L-\displaystyle\frac{3iK^\pm}{8}
\displaystyle\sum_{i,j=1}^\Lambda C_{ij}^\pm p_i\right.\right.\\[5mm]
&&\left.\left.-\displaystyle\frac{(K^{\pm})^2}{4}
\displaystyle\sum_{i,j,k=1}^\Lambda
C_{ij}^\pm C_{kj}^\pm p_ip_k\right)
\right]dp+O(\displaystyle\frac{1}{\l^2})\\[5mm]
&=&c^\pm\left(\displaystyle\frac{\sqrt{\pi}}{2}\right)^\L
\displaystyle\int_{\Rb^{\L}}exp\left\{ -K^\pm 
\displaystyle\sum_{i,j=1}^\Lambda C_{ij}^\pm 
(p_i-i(a^\pm+1))(p_j-i(a^\pm+1))\right.\\[5mm]
&&\left.-K^\pm\displaystyle (a^\pm+1)^2
\displaystyle\sum_{i,j=1}^\Lambda C_{ij}^\pm\right\}
\left[1+\displaystyle\frac{1}{\l}\left(
\displaystyle\frac{3}{16}\L-\displaystyle\frac{3iK^\pm}{8}
\displaystyle\sum_{i,j=1}^\Lambda C_{ij}^\pm p_i\right.\right.\\[5mm]
&&\left.\left.-\displaystyle\frac{(K^{\pm})^2}{4}
\displaystyle\sum_{i,j,k=1}^\Lambda
C_{ij}^\pm C_{kj}^\pm p_ip_k\right)
\right]dp+O(\displaystyle\frac{1}{\l^2}),
\ea
$$
where it has been taken into account that
$$
\displaystyle\int_{\Rb^{\L}}\int_{\Rb^{\L}}O_{p,y}({1\over\l^2})dpdy
=O({1\over\l^2}),
$$
$O_{p,y}(\displaystyle\frac{1}{\l^2})$ standing for the remainder in the
expansion of the integrand in (\ref{17}) up to order
$\displaystyle\frac{1}{\l^2}$.

Let $p_i\to p_i+i(a^\pm+1)$.
$$
\ba{rcl}
Z_\L^\pm
&=&c^\pm\left(\displaystyle\frac{\sqrt{\pi}}{2}\right)^\L
exp\left\{-K^\pm\displaystyle (a^\pm+1)^2\displaystyle
\sum_{i,j=1}^\Lambda C_{ij}^\pm\right\}
\displaystyle\int_{\Rb^{\L}}exp\left\{ -K^\pm 
\displaystyle\sum_{i,j=1}^\Lambda C_{ij}^\pm p_ip_j\right\}\\[5mm] 
&&\left[1+\displaystyle\frac{1}{\l}\left(
\displaystyle\frac{3}{16}\L-\displaystyle\frac{3iK^\pm}{8}
\displaystyle\sum_{i,j=1}^\Lambda C_{ij}^\pm(p_i+i(a^\pm+1))
\right.\right.\\[5mm]
&&\left.\left.-\displaystyle\frac{(K^{\pm})^2}{4}\displaystyle
\sum_{i,j,k=1}^\Lambda
C_{ij}^\pm C_{kj}^\pm (p_i+i(a^\pm+1))(p_k+i(a^\pm+1))\right)
\right]dp+O(\displaystyle\frac{1}{\l^2})\\[5mm]
&=&c^\pm\left(\displaystyle\frac{\sqrt{\pi}}{2}\right)^\L
exp\left\{-K^\pm\displaystyle (a^\pm+1)^2\displaystyle
\sum_{i,j=1}^\Lambda C_{ij}^\pm\right\}
\displaystyle\int_{\Rb^{\L}}exp\left\{ -K^\pm 
\displaystyle\sum_{i,j=1}^\Lambda C_{ij}^\pm p_ip_j\right\}\\[5mm] 
&&\left[1+\displaystyle\frac{1}{\l}\left(
\displaystyle\frac{3}{16}\L+\displaystyle\frac{3K^\pm}{8}
\displaystyle\sum_{i,j=1}^\Lambda C_{ij}^\pm (a^\pm+1)
+\displaystyle\frac{(K^{\pm})^2(a^\pm+1)^2}{4}
\displaystyle\sum_{i,j,k=1}^\Lambda
C_{ij}^\pm C_{kj}^\pm\right.\right.\\[5mm]
&&\left.\left.
-\displaystyle\frac{(K^{\pm})^2}{4}\displaystyle\sum_{i,j,k=1}^\Lambda
C_{ij}^\pm C_{kj}^\pm p_ip_k\right)
\right]dp+O(\displaystyle\frac{1}{\l^2}),
\ea
$$
where
$$
\ba{l}
\displaystyle\int_{\Rb^{\L}}exp\left\{ -K^\pm 
\displaystyle\sum_{i,j=1}^\Lambda C_{ij}^\pm p_ip_j\right\}
\displaystyle\sum_{i,j,k=1}^\Lambda
C_{ij}^\pm C_{kj}^\pm p_ip_k dp\\[5mm]
=\displaystyle\int_{\Rb^{\L}}exp\left\{ -K^\pm 
\displaystyle\sum_{i=1}^\L\l_i^\pm q_i^2\right\}
\displaystyle\sum_{i,j,m,n=1}^\Lambda
((C^{\pm})^2)_{ij} A_{im}A_{jn}q_m q_ndq\\[5mm]
=\displaystyle\int_{\Rb^{\L}}exp\left\{ -K^\pm 
\displaystyle\sum_{i=1}^\L\l_i^\pm q_i^2\right\}
\displaystyle\sum_{i=1}^\Lambda
(\t{A}(C^\pm)^2 A)_{ii} q_i^2dq\\[5mm]
=\displaystyle\int_{\Rb^{\L}}exp\left\{ -K^\pm 
\displaystyle\sum_{i=1}^\L\l_i^\pm q_i^2\right\}
\displaystyle\sum_{i=1}^\Lambda
(\l_i^\pm)^2 q_i^2dq
=\displaystyle\frac{1}{2K^\pm}\left(
\displaystyle\frac{\pi}{K^\pm}\right)^{\frac{\L}{2}}
\displaystyle\frac{1}{\sqrt{detC^\pm}}TrC^\pm.
\ea
$$

Therefore we get
\be\label{45}
\ba{rcl}
Z_\L^\pm &=&exp\left\{\pm\l^\pm K^\pm\L-K^\pm(2a^\pm +1)
\displaystyle\sum_{i,j=1}^\Lambda C_{ij}^\pm\right\}
\left[1+\displaystyle\frac{1}{\l}\left(
\displaystyle\frac{3}{16}\L+\displaystyle\frac{3K^\pm}{8}
\displaystyle\sum_{i,j=1}^\Lambda C_{ij}^\pm(a^\pm+1)\right.\right.\\[5mm]
&&\left.\left.+\displaystyle\frac{(K^{\pm})^2(a^\pm+1)^2}{4}
\displaystyle\sum_{i,j,k=1}^\Lambda
C_{ij}^\pm C_{kj}^\pm
-\displaystyle\frac{K^\pm}{8}TrC^\pm\right)
\right]+O(\displaystyle\frac{1}{\l^2}).
\ea
\ee

\newpage
Using (\ref{7}) and (\ref{8}) we obtain
\be\label{46}
\ba{rcl}
Z_\L^\pm &=&exp\left\{(\pm K^\pm L+K^\p)\L\right\}
\left[1+\displaystyle\frac{\L}{\l}\left(
\displaystyle\frac{3}{16}+\displaystyle\frac{3}{16}
(2K^\pm(\pm\l^\pm \mp L)-K^\p)\right.\right.\\[5mm]
&&\left.\left.+\displaystyle\frac{1}{16}(2K^\pm(\pm\l^\pm\mp L)-K^\p)^2
\mp\displaystyle\frac{K^\pm}{8}\l^\pm\right)
\right]+O(\displaystyle\frac{1}{\l^2}).
\ea
\ee
\hfill $\rule{2mm}{2mm}$

From (\ref{46}) one gets a corresponding asymptotic representation of free
energy of the system,
\be\label{47}
\ba{rcl}
F_\L^\pm &=&-k T\displaystyle\frac{1}{\L}\log Z_\L
=-k T(\pm K^\pm L+K^\p)\\[4mm]
&&-k T\displaystyle\frac{1}{\L}\log
\left[1+\displaystyle\frac{\L}{\l}\left(
\displaystyle\frac{3}{16}+\displaystyle\frac{3}{16}
(2K^\pm(\pm\l^\pm \mp L)-K^\p)\right.\right.\\[5mm]
&&\left.\left.+\displaystyle\frac{1}{16}(2K^\pm(\pm\l^\pm\mp L)-K^\p)^2
\mp\displaystyle\frac{K^\pm}{8}\l^\pm\right)
+O(\displaystyle\frac{1}{\l^2})\right].
\ea
\ee

$\underline{\bf Conclusions}$
From the integral representation of the partition function
for general $n$-dimensional Ising models with both nearest and non-nearest
neighbours interactions we have proved that for given $K$, $K^\p$, 
the partition function of n-dimensional homogeneous systems
on a square lattice is uniquely given by the eigenvalues of the
related interaction coupling matrix. 
For one-dimensional homogeneous ferromagnetic systems
with positive coupling coefficients, the partition function 
satisfies a special equality which means that the variation of the interaction
couplings of a system is related to the variations of 
the external magnetic field and the temperature. 
For some special cases of interaction coupling, we obtained, for a Ising model
in $n$-dimensions, an exact
solution for the partition function in terms of Bessel functions.
We also calculated the partition function of the $n$-dimensional Ising
model to order $O(1/\l^2)$.
It turned out that the leading terms of
the partition function for any homogeneous systems in arbitrary
dimensions
are given by the largest positive eigenvalue (resp. the largest absolute
value of the negative eigenvalues)
of the related interaction coupling matrix for ferromagnetic (resp.
antiferromagnetic) systems. 

The advantage of our integration approach
is that one can analyse the partition function of the Ising model
for various interaction couplings on lattices of arbitrary dimensions
in terms of integrals.
More results could be obtained by studying the properties of other
interaction coupling matrices. It would also be interesting to study the
relation between the largest eigenvalues of the interaction coupling
matrices in our integration approach and the largest eigenvalues of
the usual transfer matrix approach to Ising models, 
see e.g., \cite{onsager,baxter,montroll}. Moreover
the integral representation formula can also serve as an alternative
way to study the correlation functions, and as a mean to provide
numerical approximations to the thermodynamic functions of the Ising
model.

\vspace{4ex}
ACKNOWLEDGEMENTS: The original idea for the integral representation was
formed in discussion of the first author with the late Raphael H\o egh-
Krohn, in the fall of '87. We dedicate this work to his beloved memory.

\end{document}